\documentclass{article}%
\usepackage{amsmath}
\usepackage{amsfonts}
\usepackage{amssymb}
\usepackage{graphicx}
\usepackage{color}
\usepackage{multirow}
\usepackage{bbm}
\usepackage{float}%
\providecommand{\U}[1]{\protect\rule{.1in}{.1in}}
\graphicspath{{./img/}}
\newtheorem{thm}{Theorem}

\newtheorem{cor}[thm]{Corollary}

\newtheorem{prop}[thm]{Proposition}

\newcommand{\defeq}{\stackrel{\text{def}}{=}}
\begin{document}

\title{Semi-tractability of optimal stopping problems via a weighted stochastic mesh algorithm}
\author{D. Belomestny, M. Kaledin and J. Schoenmakers}
\maketitle

\begin{abstract}
In this article we propose a Weighted Stochastic Mesh (WSM) Algorithm  for approximating the value of a discrete and continuous
time optimal stopping problem. We prove that in the discrete case the WSM algorithm leads to
 semi-tractability of the corresponding optimal problems  in the sense that its complexity is
bounded in order by  $\varepsilon^{-4}\log^{d+2}(1/\varepsilon)$ with $d$ being
the dimension of the underlying Markov chain. Furthermore we study the WSM
approach in the context of continuous time optimal stopping problems and
derive the corresponding complexity bounds. Although we can not prove semi-tractability  in this case, our bounds turn out to be the tightest
ones among the bounds known for the existing algorithms in the literature. We
illustrate our theoretical findings by a numerical example.

\end{abstract}

\section{Introduction}

The theory of optimal stopping is concerned with the problem of choosing a
time to take a particular action, in order to maximize an expected reward or
minimize an expected cost. Such problems can be found in many areas of
statistics, economics, and mathematical finance (e.g. the pricing problem of
American options). Primal and dual approaches have been developed in the
literature giving rise to Monte Carlo algorithms for high-dimensional discrete
time stopping problems. Solving high-dimensional discrete optimal stopping
problems is usually based on a backward dynamic programming principle which is
in some sense contradictory to the forward nature of Monte Carlo simulation.
Much research was focused on the development of fast methods to compute
approximations to the optimal value function. Most of these methods are based
on some type of regression on Monte Carlo paths, see \cite{belomestny2018advanced} for an overview.  One of the most widely
adopted regression algorithms by practitioners is the Longstaff-Schwartz 
algorithm. It is based on approximating conditional expectations by
least-squares regression on a given basis of functions. Longstaff and
Schwartz~\cite{J_LS2001} demonstrated the efficiency of their least-squares
approach through a number of numerical examples, and in \cite{CLP} and
\cite{Z} general convergence properties of the method were established. In
particular, it follows from Corollary~3.10 in \cite{Z} that for a fixed number
$L$ of stopping opportunities and a popular choice of polynomial basis
functions of degree less or equal to $m$, the error of estimating the
corresponding value function at one point is of order
\begin{equation}
5^{L}\left(  \sqrt{\frac{m^{d}}{N}}+\frac{1}{m^{\alpha}}\right)  ,\label{lstv}%
\end{equation}
where $N$ is the number of paths used to perform regression, $\alpha\geq1$ is
related to smoothness of the corresponding conditional expectation operator,
$d$ is dimension of the underlying state space. On the other hand, the
computational cost of the least-squares MC algorithm is of order $Nm^{2d}L$
due to the computation of a (random) pseudo-inverse at every stopping date.
After balancing the variance and the approximation errors in (\ref{lstv}), one
obtains that complexity of the least-squares approach, that is, the (minimal)
number of \textquotedblleft elementary\textquotedblright\ evaluations needed
to construct an approximation for the value function with accuracy
$\varepsilon,$ is bounded up to a constant not depending on $L$ by
\begin{equation}
\mathcal{C}_{L}\left(  \varepsilon,d\right)  =\frac{L\,5^{L(2+3d/\alpha)}%
}{\varepsilon^{2+3d/\alpha}}.\label{CL}%
\end{equation}
This implies
\begin{eqnarray}
\label{eq:ls-semitract}
\limsup_{d\nearrow\infty}\limsup_{\varepsilon\searrow0}\frac{\log
\mathcal{C}_{L}\left(  \varepsilon,d\right)  }{d\log(\varepsilon^{-1}%
)}=3/\alpha.
\end{eqnarray}
Furthermore, if we next want to construct an approximation for a continuous time
optimal stopping problem, then we need to let $L\rightarrow\infty$ resulting
in the complexity bound
\[
\mathcal{C}_{\infty}(\varepsilon,d)=O\left(  \frac{\varepsilon^{-1/\beta
}\,5^{(2+3d/\alpha)\varepsilon^{-1/\beta}}}{\varepsilon^{2+3d/\alpha}}\right)
,
\]
where it is assumed that the error due to the time discretization is of order
$L^{-\beta}$ for some $0<\beta<1,$ independent of $d.$ This implies that%
\[
\lim_{\varepsilon\searrow0}\frac{\log\mathcal{C}_{\infty}(\varepsilon,d)}%
{\log(1/\varepsilon)}=\infty,
\]
showing that complexity of the least squares algorithms for continuous optimal
stopping problems may even grow faster than $\exp(1/\varepsilon)$. Similar complexity bounds can be derived for other simulation based approximation algorithms, see \cite{goldberg2018beating} for a novel nested type MC approach with complexity  depending polynomially on \(d\) and exponentially in  $1/\varepsilon.$
\par
We call a problem \textit{semi-tractable} if there is an algorithm to solve it with
complexity $\mathcal{C}(\varepsilon,d)$ satisfying
\begin{eqnarray}
\label{eq:weak-tract}
\lim_{d\nearrow\infty}\lim_{\varepsilon
\searrow0}\frac{\log\mathcal{C}\left(  \varepsilon,d\right)  }{d\log
(1/\varepsilon)}=0.
\end{eqnarray}
 Our definition of
tractability should be contrasted to the definition in \cite{NoWo} where a
problem is said to be (weakly) tractable, if there is an algorithm to solve it with
complexity $\mathcal{C}(\varepsilon,d)$ satisfying
\[
\lim_{d+\varepsilon^{-1}\nearrow\infty}\frac{\log\mathcal{C}\left(
\varepsilon,d\right)  }{d+\varepsilon^{-1}}=0.
\]
This definition seems to be counter-intuitive as it renders a problem
with, for example, an algorithmic complexity of order $d^{2}\exp(1/\left(
\varepsilon\log\log...\log\varepsilon^{-1}\right)  )$ to be (weakly) tractable
while an algorithm with complexity $2^{d}/\varepsilon$ is not. In our setting
the dimension $d$ is typically fixed and the complexity rate with respect to
$\varepsilon$ is of primary importance. In this paper we show that the
discrete time optimal stopping problems are semi-tractable in the sense
of \eqref{eq:weak-tract}. To this end we revisit the mesh method of Broadie and
Glasserman~\cite{J_BrGl}. By enhancing it with a suitable regularisation, we
prove that under mild conditions, the complexity of the resulting
WSM (Weighted Stochastic Mesh) algorithm satisfies \eqref{eq:weak-tract},
provided the transition densities of the underlying Markov chain are
analytically known or can be well approximated. Our algorithm bears some
similarity to the random grid algorithm of Rust~\cite{rust1997using}. However,
Rust~\cite{rust1997using} studied the Markovian decision problems in discrete
time with compact state space. Let us also remark that a complete convergence
as well as complexity analysis of the mesh method is still missing in the
literature, for some preliminary results see Agarwal and
Juneja~\cite{agarwal2013comparing}. In the case of continuous time optimal
stopping problems we need not to assume that the transition densities are
known but can use the Gaussian transition densities of the corresponding Euler
scheme. This results in an algorithm which has complexity of order
$O(c^{d}\varepsilon^{-(2d+14)})$ for some constant $c>1.$ Although this does not imply  semi-tractability of continuous time optimal stopping problems, the proposed algorithm is very simple and its complexity remains provably polynomial in \(\varepsilon\) as opposite to the least squares approaches. To compare different algorithms for continuous time optimal stopping problems, we introduce the so-called \textit{semi-tractability} index
\begin{eqnarray}
\label{eq:weak-tractindex}
\Gamma\defeq\limsup_{d\nearrow\infty}\limsup_{\varepsilon\searrow 0}\frac{\log\mathcal{C}\left(  \varepsilon,d\right)  }{d\log(1/\varepsilon)}.
\end{eqnarray}
It turns out that the WSM algorithm has the smallest  semi-tractability index among existing algorithms for continuous time optimal stopping problems.
\par
The paper is organized as follows. A description of the proposed algorithm is
given in Section~\ref{sec:alg}. Section~\ref{sec:conv} is devoted to
convergence and complexity analysis of our algorithm. In
Section~\ref{sec:cont} we turn to continuous time optimal stopping problems.
All proofs are collected in Section~\ref{sec:proofs}.

\section{Discrete time optimal stopping problems}

\label{sec:alg} We begin with the description of the WSM algorithm
 for discrete time optimal stopping problems. Let us assume a finite set
of stopping dates $\left\{  0,\ldots,L\right\}  ,$ for some natural $L>0,$ and
let $(Z_{l},$ $l=0,\ldots,L)$ be a Markov chain in $\mathbb{R}^{d},$ adapted
to a filtration $\left(  \mathcal{F}_{l},\,l=0,\ldots,L\right)  .$ For a given
set of nonnegative reward functions $g_{l},$ $l=0,\ldots,L,$ on $\mathbb{R}%
^{d},$ we then consider the discrete Snell envelope process:
\begin{equation}
U_{l}=U_{l}(Z_{l})\overset{\text{def}}{=}\operatornamewithlimits{esssup}_{\tau
\in\mathcal{T}_{l,L}}\mathsf{E}_{l}\left[  g_{\tau}(Z_{\tau})\right]
,\label{disc}%
\end{equation}
where $\mathcal{T}_{l,L}$ stands for the set of $\mathcal{F}$-stopping times
with values in the set $\{l,\ldots,L\},$ and $\mathsf{E}_{l}:=\mathsf{E}%
_{\mathcal{F}_{l}}$ stands for the $\mathcal{F}_{l}$-conditional expectation,
and the measurable functions $U_{l}(\cdot)$ exist due to Markovianity of the
process $(Z_{l})_{l\geq0}.$

For simplicity and without loss of generality we assume that the Markov chain
$(Z_{l})_{l\geq0}$ is time homogeneous with $l$-steps transition density
denoted by $p_{l}(y|x)$ and one-step density denoted by $p(y|x)=p_{1}(y|x),$
so that
\[
\mathbb{P}\left[  \left.  Z_{k+1}\in dy\right\vert Z_{k}=x\right]  =p(y|x)dy.
\]
Fix some $x_{0}\in\mathbb{R}^{d}$ and assume that $Z_{0}=x_{0}.$ It is well
known that the Snell envelope (\ref{disc}) satisfies the dynamic program
principle,
\begin{align}
&  U_{L}(Z_{L})=g_{L}(Z_{L}),\label{eq:ubar_dp-1}\\
&  U_{l}(Z_{l})=\max\left\{  g_{l}(Z_{l}),\mathsf{E}\left[  \left.
U_{l+1}(Z_{l+1})\right\vert Z_{l}\right]  \right\}  ,\quad l=0,\ldots
,L-1.\nonumber
\end{align}
Next we fix some $R>0$ and define a truncated version of the above dynamic
program via
\begin{align}
&  \widetilde{U}_{L}(Z_{L})=g_{L}(Z_{L})\cdot\mathbbm{1}_{Z_{L}\in B_{R}%
},\label{BDPt}\\
&  \widetilde{U}_{l}(Z_{l})=\max\left\{  g_{l}(Z_{l}),\mathsf{E}\left[
\left.  \widetilde{U}_{l+1}(Z_{l+1})\right\vert Z_{l}\right]  \right\}
\cdot\mathbbm{1}_{Z_{l}\in B_{R}},\quad l=0,\ldots,L-1,\nonumber
\end{align}
where $B_{R}\overset{\text{def}}{=}\left\{  z:\left\vert z-x_{0}\right\vert
\leq R\right\}  .$ Thus, by construction, $\widetilde{U}_{l}$ vanishes outside
the ball $B_{R}.$ Also by construction it holds that%
\begin{equation}
\Vert\widetilde{U}_{l}\Vert_{\infty}\leq G_{R}\overset{\text{def}}{=}%
\max_{0\leq l\leq L}\sup_{z\in B_{R}}g_{l}(z),\label{boundG}%
\end{equation}
which is easily seen by backward induction. In view of (\ref{BDPt}) we may
write
\[
\mathsf{E}\left[  \left.  \widetilde{U}_{l+1}(Z_{l+1})\right\vert
Z_{l}=x\right]  =\int\widetilde{U}_{l+1}(y)\frac{p(y|x)}{p_{l+1}(y|x_{0}%
)\,}p_{l+1}(y|x_{0})\,dy.
\]
Now assume that we have a set of trajectories $Z_{l}^{(n)},$ $l=0,\ldots,L,$
with $Z_{0}^{(n)}=x_{0},$ $n=1,\ldots,N,$ simulated according to the one-step
transition density $p,$ and consider the approximation:
\[
\mathsf{E}\left[  \left.  \widetilde{U}_{l+1}(Z_{l+1})\right\vert
Z_{l}=x\right]  \approx\frac{1}{N}\sum_{n=1}^{N}\widetilde{U}_{l+1}%
(Z_{l+1}^{(n)})\frac{p(Z_{l+1}^{(n)}|x)}{p_{l+1}(Z_{l+1}^{(n)}|x_{0})},
\]
where in view of the Chapman-Kolmogorov equation
\[
p_{l+1}(Z_{l+1}^{(n)}|x_{0})=\int p(Z_{l+1}^{(n)}|z)p_{l}(z|x_{0}%
)\,dz\approx\frac{1}{N}\sum_{m=1}^{N}p(Z_{l+1}^{(n)}|Z_{l}^{(m)}).
\]
Hence we have approximately
\begin{equation}
\mathsf{E}\left[  \left.  \widetilde{U}_{l+1}(Z_{l+1})\right\vert
Z_{l}=x\right]  \approx\sum_{n=1}^{N}\widetilde{U}_{l+1}(Z_{l+1}^{(n)}%
)\frac{p(Z_{l+1}^{(n)}|x)}{\sum_{m=1}^{N}p(Z_{l+1}^{(n)}|Z_{l}^{(m)}%
)}.\label{app}%
\end{equation}
We thus propose the following algorithm. We start with
\[
\overline{U}_{L}(Z_{L}^{(n)})\overset{\text{def}}{=}g_{L}(Z_{L}^{(n)}%
)\mathbbm{1}_{Z_{L}^{(n)}\in B_{R}}%
\]
for $n=1,\ldots,N.$ Once $\overline{U}_{l+1}$ is constructed on the grid for
$0<l+1\leq L,$ we set
\begin{equation}
\overline{U}_{l}(Z_{l}^{(r)})\overset{\text{def}}{=}\max\left\{  g_{l}%
(Z_{l}^{(r)}),\sum_{n=1}^{N}\overline{U}_{l+1}^{(n)}(Z_{l+1}^{(n)}%
)\frac{p(Z_{l+1}^{(n)}|Z_{l}^{(r)})}{\sum_{m=1}^{N}p(Z_{l+1}^{(n)}|Z_{l}%
^{(m)})}\right\}  \mathbbm{1}_{Z_{l}^{(r)}\in B_{R}},\label{ABDP}%
\end{equation}
for $r=1,\ldots,N.$ By construction, each function $\overline{U}_{l}$ vanishes
outside the ball $B_{R}.$ Working all the way down to $l=0$ results in the
approximation:
\[
\overline{U}_{0}=\max\left[  g_{0}(x_{0}),\sum_{n=1}^{N}\overline{U}_{1}%
^{(n)}(Z_{1}^{(n)})\frac{p(Z_{1}^{(n)}|x_{0})}{\sum_{m=1}^{N}p(Z_{1}%
^{(n)}|x_{0})}\right]
\]
for $U_{0}.$ As such the presented algorithm is closely related to the mesh
method of Broadie and Glasserman~\cite{J_BrGl} apart from truncation at level
$R$ and a special choice of weights.

\subsection{Cost estimation}

Let us estimate the cost of carrying out the backward dynamic program
(\ref{ABDP}). One needs to compute $p(Z_{l+1}^{(n)}|Z_{l}^{(m)})$ for all
$l=1,\ldots,L,$ $n,$ $m=1,\ldots,N.$ This can be done at a cost of order
$N^{2}Lc_{f}^{(d)},$ where $c_{f}^{(d)}$ is the cost of evaluating a (typical)
function of $2d$ arguments. In the typical situation $c_{f}^{(d)}$ is
proportional to $d.$ The evaluation of
\[
\frac{1}{N}\sum_{m=1}^{N}p(Z_{l+1}^{(n)}|Z_{l}^{(m)})
\]
for $l=1,...,L,$ $n=1,...,N,$ has a cost of order $N^{2}Lc_{\ast}$ with
$c_{\ast}$ being the cost of an elementary numerical operation, which is
negligible if $c_{\ast}\ll c_{f}^{(d)}.$ So the overall cost of carrying out
the backward dynamic program (\ref{ABDP}) is of order $N^{2}Lc_{f}^{(d)}.$

\subsection{Error and complexity analysis}

\label{sec:conv} In this section we analyze  convergence of the  WSM
estimate  (\ref{ABDP}) to the solution of the discrete optimal stopping
problem (\ref{disc}) for $l=0$ and a fixed $x_{0}\in\mathbb{R}^{d}$ as \(N\to \infty.\)
Let us  first bound a distance between $U_{l}$ and $\widetilde{U}_{l},$
$l=0,\ldots,L.$

\begin{prop}
\label{eq:u-ut} With%
\[
\varepsilon_{l,R}\overset{\text{def}}{=}\int_{\left\vert x-x_{0}\right\vert
>R}U_{l}(x)p_{l}(x|x_{0})\,dx
\]
$l=0,\ldots,L,$ it holds that
\begin{equation}
\int\bigl |U_{l}(x)-\widetilde{U}_{l}(x)\bigr |p_{l}(x|x_{0})\,dx\leq
\sum_{j=l}^{L}\varepsilon_{j,R}. \label{es}%
\end{equation}

\end{prop}

\begin{prop}
\label{corL} Suppose that%
\begin{align}
\max_{0\leq l\leq L}g_{l}(x)\leq c_{g}(1+\left\vert x\right\vert ), \quad
x\in\mathbb{R}^{d}\label{eq:assg}%
\end{align}
and that%
\begin{align}
\label{eq:assz}\mathsf{E}\left[  \left.  \max_{l\leq l^{\prime}\leq
L}\left\vert Z_{l^{\prime}}\right\vert \right\vert Z_{l}=x\right]  \leq
c_{Z}(1+\left\vert x\right\vert ), \quad x\in\mathbb{R}^{d}.
\end{align}
Suppose further that for some $\varkappa,$ $\alpha>0,$ and $l=1,\ldots,L,$
\begin{align}
p_{l}(y|x)\leq\frac{\varkappa}{(2\pi\alpha l)^{d/2}}e^{\frac{|x-y|^{2}%
}{2\alpha l}}.\label{eq:assp}%
\end{align}
for all $x,y\in\mathbb{R}^{d}.$ One then has%
\begin{multline}
\int\bigl |U_{l}(x)-\widetilde{U}_{l}(x)\bigr |p_{l}(x|x_{0})\,dx\label{es2}\\
\leq Lc_{g}\varkappa\left(  1+c_{Z}+c_{Z}\left\vert x_{0}\right\vert
+c_{Z}\sqrt{d\alpha L}\right)  2^{d/4}e^{-\frac{R^{2}}{8\alpha L}}.
\end{multline}
\end{prop}
Next we control the discrepancy between \(\overline{U}_{0}\) and \(\widetilde{U}_{0}.\)
\begin{prop}
\label{thm: main-err} With
\begin{equation}
F_{R}^{2}\overset{\text{def}}{=}\int\int_{|y-x_{0}|\leq R}\frac{p^{2}%
(y|x)}{p_{l+1}(y|x_{0})}\,p_{l}(x|x_{0})\,dxdy, \label{Fr}%
\end{equation}
and $N$ such that $\left(  1+F_{R}\right)  /\sqrt{N}<1,$ it holds that%
\[
\mathsf{E}\left[  \bigl |\overline{U}_{0}-\widetilde{U}_{0}\bigr |\right]
\leq\left(  3+\sqrt{2}\right)  LG_{R}\frac{1+F_{R}}{\sqrt{N}}.
\]

\end{prop}

\begin{cor}
\label{corL1} Under the assumptions of Proposition~\ref{corL}, we have for
(\ref{Fr}) the estimate
\[
F_{R}^{2}\leq\frac{\varkappa}{(2\pi\alpha)^{d/2}}\mathtt{Vol}(B_{R}%
)=\frac{\varkappa R^{d}}{(2\alpha)^{d/2}\Gamma\left(  1+d/2\right)  }%
\leq\varkappa\left(  e/\alpha\right)  ^{d/2}R^{d}d^{-d/2},
\]
where the last inequality follows from $\Gamma\left(  1+a\right)  \geq
a^{a}e^{-a}$ for any $a\geq1/2.$ Then by combining (\ref{es2}) with
Proposition~\ref{thm: main-err} we obtain the error estimate,%
\begin{multline}
\mathsf{E}\left[  \bigl |U_{0}-\overline{U}_{0}\bigr |\right]  \leq
Lc_{g}\varkappa\left(  1+c_{Z}+c_{Z}\left\vert x_{0}\right\vert +c_{Z}%
\sqrt{d\alpha L}\right)  2^{d/4}e^{-\frac{R^{2}}{8\alpha L}}\label{errL}\\
+\left(  3+\sqrt{2}\right)  Lc_{g}(1+R)\frac{1+\varkappa^{1/2}\left(
e/\alpha\right)  ^{d/4}R^{d/2}d^{-d/4}}{\sqrt{N}}.
\end{multline}

\end{cor}

\begin{prop}
\label{compl-wsm-discr}
Under the assumptions of Proposition~\ref{corL} the complexity of the WSM
algorithm is bounded from above by%
\begin{multline}
\mathcal{C}(\varepsilon,d)=c_{1}\alpha^{2}c_{g}^{4}\varkappa^{2}c_{f}%
^{(d)}c_{2}^{d}L^{d+7}\varepsilon^{-4}\\
\times\log^{d+2}\left[  \frac{L\left(  1+c_{Z}+c_{Z}\left\vert x_{0}%
\right\vert \right)  e^{\frac{c_{Z}\sqrt{\alpha L}}{1+c_{Z}+c_{Z}\left\vert
x_{0}\right\vert }}2^{3/4}\left(  c_{g}\varkappa\vee1\right)  }{\varepsilon
}\right]  ,\label{comLL}%
\end{multline}
where $c_{1}>0$ and $c_{2}>1$ are natural constants and $c_{f}^{(d)}$ stands
for the cost of computing the transition density $p_{l}(y|x)$ at one point
$(x,y).$
\end{prop}

\begin{cor}
For a fixed $L>0$ the discrete time optimal stopping problem \eqref{disc} with
$g$ and $(Z_{l})_{l\geq0}$ satisfying \eqref{eq:assg}, \eqref{eq:assz} and
\eqref{eq:assp} is semi-tractable, provided that the complexity of computing the transition
density $p_{l}(y|x)$ at one point $(x,y)$ is at most polynomial in $d.$ Different approximation algorithms for discrete time optimal stopping problems can be compared using the semi-tractability index \eqref{eq:weak-tractindex}.  For example, it follows from \eqref{eq:ls-semitract}  that the semi-tractability index of the least-squares (LS) approach is equal to \(3/\alpha.\) Hence it tends to \(0\) as the smoothness of the problem increases. Moreover from inspection of  Theorem~2.4  in \cite{Bally}, we see that the Quantisation Tree (QT) method  has semi-tractability index \(2.\)
\begin{table}
\begin{center}
\begin{tabular}{ |c|c|c| }
 \hline
 LS & WSM & QTM  \\
 \hline
\(3/\alpha\) & \(0\) &  \(2\) \\
 \hline
\end{tabular}
\end{center}
\caption{Semi-tractability index \(\Gamma\) of different algorithms for discrete time optimal stopping problems \label{tab:disct}}
\end{table}
\end{cor}

\subsection{Approximation of the transition density}

\label{sec:apprdens} A crucial condition for semi-tractability to hold is
availability of the transition density $p(y|x)$ of the chain $(Z_{l})_{l\geq
0}$ in closed form. However it can be shown that if a sequence of
approximating densities $p^{n}(y|x),$ $n\in\mathbb{N},$ converging to $p(y|x)$
can be constructed in such a way that
\begin{equation}
\left\vert \frac{p^{n}(y|z)-p(y|z)}{p^{n}(y|z)}\right\vert \lesssim
\frac{\left(  1+|y-x_{0}|^{m}+|z-x_{0}|^{m}\right)  ^{n}}{n!},\quad y,z\in
B_{R_{n}} \label{eq:ass-dens}%
\end{equation}
for some $m\in\mathbb{N}$ and a sequence $R_{n}\nearrow\infty,$ $n\nearrow
\infty,$ then under proper assumptions on the growth of $R_{n}$ and the cost
of computing $p^{n}$ (in fact it should be at most polynomial in $d$), one can
derive a complexity bound \(\mathcal{C}(\varepsilon,d)\) satisfying
\[
\lim_{\varepsilon\searrow0}\frac{\log\mathcal{C}(\varepsilon,d)}{\log\frac
{1}{\varepsilon}}\mbox{ is finite and does not depend on \(d\) }.
\]

To construct a sequence of  approximations $p^{n}(y|z)$ satisfying the assumption
\eqref{eq:ass-dens}, one can use various small-time expansions for transition
densities of stochastic processes, see, for example, \cite{Azencott} and
\cite{Li}. Let us exemplify this type of approximation in the case of
one-dimensional diffusion processes of the form:
\[
dX_{t}=b(X_{t})\,dt+\sigma(X_{t})\,dW_{t},\quad X_{0}=x_{0},
\]
where $b$ is a bounded function, twice continuously differentiable, with
bounded derivatives and $\sigma$ is a function with three continuous and
bounded derivatives such that there exist two positive constants
$\sigma_{\circ},\sigma^{\circ}$ with $\sigma_{\circ}\leq\sigma(x)\leq
\sigma^{\circ}.$ Consider a Markov chain $(Z_{l})_{l\geq0}$ defined as a time
discretization of $(X_{t})_{t\geq0},$ that is, $Z_{l}\overset{\text{def}}{=}
X_{\Delta l},$ $l=0,1,2,\ldots$ for some $\Delta>0.$ Under the above
conditions the following representation for the (one-step) transition density \(p\) of the chain \(Z\) is proved in \cite{Florens-Zmirou}
(see also \cite{MR872464} for more general setting):
\begin{align*}
p(y|x)=\frac{1}{\sqrt{2\pi\Delta}}\frac{1}{\sigma(y)}\exp\left(
-\frac{(s(x)-s(y))^{2}}{2\Delta}\right)  U_{\Delta}(s(x),s(y)),\quad
x,y\in\mathbb{R},
\end{align*}
with $U_{\Delta}(x,y)=R_{\Delta}(x,y)\exp\left[  \int_{0}^{x} \bar
b(z)\,dz-\int_{0}^{y} \bar b(z)\,dz\right]  ,$
\begin{align}
\label{eq:Rh}R_{\Delta}(x,y)=\mathsf{E}\left[  \exp\left(  -\Delta\int_{0}%
^{1}\bar\rho(x+z(y-x)+\sqrt{\Delta} B_{z})\,dz\right)  \right]  ,
\end{align}
where $B_{z}$ is a standard Brownian bridge, $s(x)=\int_{0}^{x} \frac
{dy}{\sigma(y)},$ $g=s^{-1}$ and
\begin{align*}
\bar\rho=(\bar b^{2}+\bar b^{\prime})/2 \quad\mbox{ with } \quad\bar b=
(b/\sigma)\circ g-\sigma^{\prime}\circ g/2.
\end{align*}
By expanding the exponent in \eqref{eq:Rh} into Taylor series, we get for
$\Delta$ small enough
\begin{multline*}
p(x|y)=\frac{1}{\sqrt{2\pi\Delta}}\frac{1}{\sigma(y)}\exp\left(
-\frac{(s(x)-s(y))^{2}}{2\Delta}\right) \\
\times\exp\left[  \int_{0}^{x} \bar b(z)\,dz-\int_{0}^{y} \bar
b(z)\,dz\right]  \sum_{k=0}^{\infty}\frac{\Delta^{k}}{k!} c_{k}(x,y)
\end{multline*}
with
\begin{align*}
c_{k}(x,y)=(-1)^{k}\mathsf{E}\left[  \left(  \int_{0}^{1}\bar\rho
(x+z(y-x)+\sqrt{\Delta} B_{z})\,dz\right)  ^{k}\right]  .
\end{align*}
If $\bar\rho$ is uniformly bounded by a constant $D>0$, then the above series
converges uniformly in $x$ and $y$ for all $\Delta$ small enough. Set
\begin{multline*}
p^{n}(x|y) =\frac{1}{\sqrt{2\pi\Delta}}\frac{1}{\sigma(y)}\exp\left(
-\frac{(s(x)-s(y))^{2}}{2\Delta}\right) \\
\times\exp\left[  \int_{0}^{x} \bar b(z)\,dz-\int_{0}^{y} \bar
b(z)\,dz\right]  \left\{  \sum_{k=0}^{n}\frac{\Delta^{k}}{k!} c_{k}(x,y)
\right\}  .
\end{multline*}
It obviously holds $p^{n}(y|x)>0 $ for $\Delta<\Delta_{0}(D)$ and
\begin{align}
\label{eq:ratio-example}\left\vert \frac{p^{n}(y|z)-p(y|z)}{p^{n}%
(y|z)}\right\vert  &  \leq\frac{(\Delta D)^{n}}{(1-\Delta D\exp(\Delta D))}%
\end{align}
uniformly for all $x,y\in\mathbb{R}.$ Hence the assumption \eqref{eq:ass-dens}
is satisfied with $m=0,$ provided that $\Delta<\Delta_{0}$ for some $\Delta
_{0}$ depending only on $D.$ Similarly if $\bar\rho\leq0,$ then
\eqref{eq:ass-dens} holds. To sample from $p^{n}$ we can use the well-known
acceptance rejection method which does not require the exact knowledge of a
scaling factor $\int p^{n}(y|x)\,dy$.

\section{Continuous time optimal stopping for diffusions}

\label{sec:cont} In this section we consider diffusion processes of the form
\begin{equation}
dX_{s}^{i}=b_{i}(X_{s})\,ds+\sum_{j=1}^{m}\sigma_{ij}(X_{s})\,dW_{s}^{j},\quad
X_{0}^{i}=x_{0}^{i},\quad i=1,\ldots,d, \label{eq: sde}%
\end{equation}
where $b:$ $\mathbb{R}^{d}\rightarrow\mathbb{R}^{d}$ and $\sigma:$
$\mathbb{R}^{d}\rightarrow\mathbb{R}^{d\times m},$ are Lipschitz continuous
and $W=(W^{1},\ldots,W^{m})$ is a $m$-dimensional standard Wiener process on a
probability space $(\Omega,\mathcal{F},P)$. As usual, the (augmented)
filtration generated by $(W_{s})_{s\geq0}$ is denoted by $(\mathcal{F}%
_{s})_{s\geq0}.$ We are interested in solving optimal stopping problems of the
form:
\begin{equation}
U_{t}^{\star}=\operatornamewithlimits{esssup}_{\tau\in\mathcal{T}_{t,T}%
}\mathsf{E}[e^{-r(\tau-t)}f(X_{\tau})|\mathcal{F}_{t}], \label{In3}%
\end{equation}
where $f$ is a given real valued function on $\mathbb{R}^{d},$ $r\geq0,$ and
$\mathcal{T}_{t,T}$ stands for the set of stopping times $\tau$ taking values
in $[t,T]$. The problem \eqref{In3} is related to the so-called free boundary
problem for the corresponding partial differential equation. Let us introduce
the  differential operator $L_{t}$ :
\[
L_{t}u(t,x)=\frac{1}{2}\sum_{i,j=1}^{d}a_{ij}(x)\frac{\partial^{2}u}{\partial
x^{i}\partial x^{j}}(t,x)+\sum_{i=1}^{d}b_{i}(x)\frac{\partial u}{\partial
x^{i}}(t,x),
\]
where
\[
a_{ij}(x)=\sum_{k=1}^{d}\sigma_{ik}(x)\sigma_{jk}(x).
\]
We denote by $X_{s}^{t,x}$ (or $X^{t,x}(s)$)$,\;s\geq T,$ the solution of
(\ref{eq: sde}) starting at  moment $t$ from $x:\;X_{t}^{t,x}=x.$ Denote by
$u(t,x)$ a regular solution of the following system of partial differential
inequalities:
\begin{gather}
\frac{\partial u}{\partial t}+L_{t}u-ru\leq0,\;u\geq f,\quad(t,x)\in
\mathbf{[}0,T\mathbf{)}\times\mathbb{R}^{d},\label{In4}\\
\left(  \frac{\partial u}{\partial t}+L_{t}u-ru\right)  (f-u)=0,\quad
(t,x)\in\mathbf{[}0,T\mathbf{)}\times\mathbb{R}^{d},\nonumber\\
u(T,x)=f(x),\quad x\in\mathbb{R}^{d},\nonumber
\end{gather}
then under some mild conditions (see, e.g. \cite{jaillet1990variational})
\begin{equation}
u(t,x)=\sup_{\tau\in\mathcal{T}_{t,T}}\mathsf{E}[e^{-r(\tau-t)}f(X_{\tau
}^{t,x})]\quad,\;(t,x)\in\mathbf{[}0,T\mathbf{]}\times\mathbb{R}^{d},
\label{In5}%
\end{equation}
that is, $u(t,x)=U_{t}^{\star}(x).$
\par
With this notation established, it is worth discussing the main issue
that we are going to address in this section. Our goal is to estimate $u(t,x)$ at a
given point $(t_{0},x_{0})$ with accuracy less than $\varepsilon$ by an
algorithm with complexity $\mathcal{C}^{\star}(\varepsilon,d)$ which is
polynomial in $1/\varepsilon$. As already mentioned in the introduction some
well known algorithms such as the regression ones fail to achieve this goal
(at least according to the existing complexity bounds in the literature). 
\par
Let us introduce the Snell envelope process:
\begin{equation}
U_{t}^{\star}\overset{\text{def}}{=}\mathrm{esssup}_{\tau\in\mathcal{T}_{t,T}%
}\mathsf{E}_{\mathcal{F}_{t}}\left[  g(\tau,X_{\tau})\right]  , \label{Snell}%
\end{equation}
where (somewhat more general than in (\ref{In3})) $g$ is a given nonnegative
function on $\mathbb{R}_{\geq0}\times\mathbb{R}^{d}.$ In the first step we
perform a time discretization by introducing a finite set of stopping dates
$t_{l}=lh,$ $l=1,\ldots,L,$ with $h=T/L$ and $L$ some natural number, and next
consider the discretized Snell envelope process:
\[
U_{t_{l}}^{\circ}(X_{t_{l}})\overset{\text{def}}{=}%
\operatornamewithlimits{esssup}_{\tau\in\mathcal{T}_{l,L}}\mathsf{E}%
_{\mathcal{F}_{t_{l}}}\left[  g(\tau,X_{\tau})\right]  ,
\]
where $\mathcal{T}_{l,L}$ stands for the set of stopping times with values in
the set $\{t_{l},\ldots,t_{L}\}.$ Note that the measurable functions $U_{t_{l}}%
^{\circ}(\cdot)$ exist due to Markovianity of the process $X.$ The error due
to the time discretization is well studied in the literature. We will rely on the
following result which is implied by Thm.~2.1 in \cite{Bally} for instance.

\begin{prop}
\label{discb} Let $g:[0,T]\times\mathbb{R}^{d}\rightarrow$ $\mathbb{R}$ be
Lipschitz continuous and $p\geq1.$ Then one has that%
\[
\max_{l=0,\ldots,L}\left\Vert U_{t_{l}}^{\star}(X_{t_{l}})-U_{t_{l}}^{\circ
}(X_{t_{l}})\right\Vert _{p}\leq\frac{c_{\circ}e^{C_{\circ}T}(1+\left\vert
x_{0}\right\vert )}{L},
\]
where the constants $c_{\circ},C_{\circ}>0$ depend on the Lipschitz constants
for $b,\sigma,$ and $g,$ respectively.
\end{prop}

In order to achieve an acceptable discretization error we choose a
sufficiently large $L,$ and then concentrate on the computation of $U^{\circ
}.$

In the next step we approximate the underlying process $X$ using some strong
discretization scheme on the time grid $t_{i}=iT/L,$ $i=0,\ldots,L,$
yielding an approximation $\overline{X}.$ It is assumed that the one step
transition densities of this scheme are explicitly known. The
simplest and the most popular scheme   is the Euler scheme,%
\begin{equation}
\overline{X}_{t_{l+1}}^{i}=\overline{X}_{t_{l}}^{i}+b_{i}(\overline{X}_{t_{l}%
})\,h+\sum_{j=1}^{m}\sigma_{ij}(\overline{X}_{t_{l}})\,\left(  W_{t_{l+1}}%
^{j}-W_{t_{l}}^{j}\right)  ,\text{ \ \ }\overline{X}_{0}^{i}=x_{0}^{i},
\label{Euler}%
\end{equation}
$i=1,\ldots,d,$ which in general has strong convergence order $1/2,$ and the
one-step transition density of the chain $(\overline{X}_{t_{l+1}})_{l\geq0}$
is given by%
\begin{align}
\bar{p}_{h}(y|x) \overset{\text{def}}{=} \frac{1}{\sqrt{\left(  2\pi h\right)
^{d}\left\vert \Sigma\right\vert }}\exp\left[  -\frac{1}{2}h^{-1}%
(y-x-b(x)h)^{\top}\Sigma^{-1}(y-x-b(x)h)\right]  \label{Edens}%
\end{align}
with $\Sigma=\sigma\sigma^{\top}\in\mathbb{R}^{d\times d}$ and \(h=T/L.\) Now we will turn
to the discrete time optimal stopping problem with possible stopping times $\{t_{l}=lh,$
$l=0,\ldots,L\}$. To
this end we introduce the discrete time Markov chain $Z_{l}\overset
{\text{def}}{=}\overline{X}_{t_{l}}$ adapted to the filtration $(\mathcal{F}%
_{l})\overset{\text{def}}{=}(\mathcal{F}_{t_{l}}),$ and $g_{l}(x)\overset
{\text{def}}{=}g(t_{l},x)$ (while abusing notation slightly) and consider the
discretized Snell envelope process
\begin{align}
\label{eq:Ul}U_{t_{l}}(\overline{X}_{t_{l}})\overset{\text{def}}%
{=}\operatornamewithlimits{esssup}_{\tau\in\mathcal{T}_{l,L}}\mathsf{E}%
_{\mathcal{F}_{t_{l}}}\left[  g(\tau,\overline{X}_{\tau})\right]
=\operatornamewithlimits{esssup}_{\iota\in\mathcal{I}_{l,L}}\mathsf{E}%
_{\mathcal{F}_{l}}\left[  g_{\iota}(Z_{\iota})\right]  \overset{\text{def}}%
{=}U_{l}(Z_{l}),
\end{align}
where $\mathcal{I}_{l,L}$ stands for the set of stopping indices with values
in  $\{l,\ldots,L\},$ and the measurable functions $U_{t_{l}}(\cdot)$ (or
$U_{l}(\cdot)$) exist due to Markovianity of the process $\overline{X}$ (or
$Z$). The distance between $U$ and $U^{\circ}$ is controlled by the next proposition.

\begin{prop}
\label{errcirc} There exists a constant $C^{\text{Euler}}>0$ depending on the
Lipschitz constants of $b,\sigma,$ and $g,$ such that
\[
\max_{l=0,...,L}\mathsf{E}\left[  \left\vert U_{t_{l}}^{\circ}(X_{t_{l}%
})-U_{t_{l}}(\overline{X}_{t_{l}})\right\vert \right]  \leq C^{\text{Euler}%
}\sqrt{h}.
\]

\end{prop}
Thus, combining Proposition~\ref{discb} and Proposition~\ref{errcirc} yields.
\begin{cor}
If $\overline{X}$ is constructed by the Euler scheme with time step size
$h=T/L,$ where $L$ is the number of discretization steps, then under the
conditions of Proposition~\ref{discb} and Proposition~\ref{errcirc} we have
that%
\begin{equation}
\mathsf{E}\left[  \left\vert U_{0}^{\star}(x_{0})-U_{0}(x_{0})\right\vert
\right]  \lesssim C^{\text{Euler}}\sqrt{h}\text{ \ \ for }h\rightarrow0,
\label{star}%
\end{equation}
where $\lesssim$ stands for inequality up to constant depending on $c_{\circ
},C_{\circ}$ and $C^{\text{Euler}}.$
\end{cor}

Since the transition densities of the Euler scheme are explicitly known (see
(\ref{Edens})), the WSM algorithm can be directly used for constructing an
approximation $\overline{U}_{0}(x_{0})$ based on the paths of the Markov chain
$(Z_{l}).$ To derive the complexity bounds of the resulting estimate, we shall make the following assumptions.

\begin{description}
\item[(AG)] Suppose that $c_{g}>0$ is such that
\begin{equation}
g(t,x)\leq c_{g}\left(  1+\left\vert x\right\vert \right)  \text{ \ \ for all
}0\leq t\leq T,\text{ }x\in\mathbb{R}^{d}. \label{ag}%
\end{equation}

\item[(AX)] Assume that there exists a constant $c_{\bar{X}}>0$ such that for
all $0\leq l\leq L,$
\begin{equation}
\mathsf{E}_{\mathcal{F}_{t_{l}}}\Bigl[  \sup_{l\leq l^{\prime}\leq
L}\left\vert \overline{X}_{l^{\prime}h}\right\vert \Big | \overline
{X}_{lh}=x\Bigr]  \leq c_{\bar{X}}\left(  1+\left\vert x\right\vert
\right),\quad x\in\mathbb{R}^{d}, \label{aX}%
\end{equation}
uniformly in  $L$ (hence $h$). This assumption is satisfied under
Lipschitz conditions on the coefficients of the SDE (\ref{eq: sde}), and can
be proved using the Burkholder-Davis-Gundy inequality and the Gronwall lemma.

\item[(AP)] Assume furthermore that $\left(  \overline{X}_{lh},\text{
}l=0,\ldots,L\right)  $ is time homogeneous with transition densities
$\overline{p}_{lh}(y|x)$ that satisfy the Aronson type inequality: there exist
positive constants $\overline{\varkappa}$ and $\overline{\alpha}$ such that
for any $x,y\in\mathbb{R}^{d}$ and any $l>0,$ it holds that
\[
\overline{p}_{lh}(y|x)\leq\frac{\overline{\varkappa}}{(2\pi\overline{\alpha
}lh)^{d/2}}e^{-\frac{|x-y|^{2}}{2\overline{\alpha}lh}}.
\]
This assumption holds if the coefficients in \eqref{eq: sde} are bounded and
$\sigma$ is uniformly elliptic.
\end{description}

The next proposition provides complexity bounds for the WSM algorithm in
the case of continuous time optimal stopping problems.

\begin{prop}
\label{corcanset} Assume that the assumptions (AG), (AX) and (AP) hold, then

\begin{itemize}
\item the cost of computing $U_{0}(x_{0})$ in \eqref{eq:Ul} for a fixed \(L>0\) with precision
$\varepsilon>0$ via the WSM algorithm is bounded above by
\begin{multline}
\mathcal{C}(\varepsilon,d)=c_{1}\overline{\alpha}^{2}c_{g}^{4}\varkappa
^{2}c_{f}^{(d)}c_{2}^{d}\frac{T^{d+7}}{h^{d+5}}\label{comc}\\
\times\varepsilon^{-4}\log^{d+2}\left[  \frac{\frac{T}{h}\left(
1+c_{\bar{X}}+c_{\bar{X}}\left\vert x_{0}\right\vert \right)
e^{\frac{c_{\bar{X}}\sqrt{\overline{\alpha}T}}{1+c_{\bar{X}%
}+c_{\bar{X}}\left\vert x_{0}\right\vert }}2^{3/4}\left(  c_{g}%
\varkappa\vee1\right)  }{\varepsilon}\right]  .
\end{multline}

\item the cost of computing $U_{0}^{\star}(x_{0})$ with an accuracy
$\varepsilon>0$ via the WSM algorithm is bounded by
\begin{multline}
\mathcal{C}^{\star}(\varepsilon,d)=c_{1}\overline{\alpha}^{2}c_{g}%
^{4}\varkappa^{2}c_{f}^{(d)}c_{2}^{d}\frac{T^{d+7}}{\varepsilon^{2d+14}%
}\label{comc-star}\\
\times\log^{d+2}\left[  \frac{T\left(  1+c_{\bar{X}}+c_{\bar{X}%
}\left\vert x_{0}\right\vert \right)  e^{\frac{c_{\bar{X}}\sqrt
{\overline{\alpha}T}}{1+c_{\bar{X}}+c_{\bar{X}}\left\vert
x_{0}\right\vert }}2^{3/4}\left(  c_{g}\varkappa\vee1\right)  }{\varepsilon
}\right]  .
\end{multline}

\end{itemize}
\end{prop}

The first statement follows directly from Proposition~\ref{compl-wsm-discr}
by taking in (\ref{comLL}),
$\alpha=\overline{\alpha}h,$ $c_{Z}=c_{\bar{X}},$ and $L=T/h.$ Then by
setting $h\asymp\varepsilon^{2}$ we obtain \eqref{comc-star} (with possibly
modified natural constants $c_{1},c_{2}$).

\paragraph{Discussion}

As can be seen from \eqref{comc-star},
\begin{equation}
\Gamma_{\text{WSM}}=\lim_{d\nearrow\infty}\lim_{\varepsilon\searrow0}\frac{\log\mathcal{C}^{\star
}(\varepsilon,d)}{d\,\log\varepsilon^{-1}}=2 \label{limq}%
\end{equation}
and this shows the efficiency of the proposed algorithm as compared to the existing
algorithms for continuous time optimal stopping problems at least as far as
the semi-tractability index is concerned. Indeed, the only algorithm available in the literature with a provably finite limit of type (\ref{limq}) is the  quantization tree  algorithm (QTA) of
Bally, Pag\`{e}s, and Printems~\cite{Bally}. Indeed, by tending the number of
stopping times and the quantization number to infinity such that the
corresponding errors in Thm.~2.4-b in \cite{Bally} are balanced, we derive the following  complexity upper bound
\begin{equation}
\mathcal{C}^{\star}_{\text{QTA}}\left(  \varepsilon,d\right)  =O\left(  \frac
{1}{\varepsilon^{6d+6}}\right)
\end{equation}
Hence \(\Gamma_\text{QTA}=6.\)
\begin{table}
\begin{center}
\begin{tabular}{ |c|c|c| }
 \hline
 LS & WSM & QTA  \\
 \hline
\(\infty\) & \(2\) &  \(6\) \\
 \hline
\end{tabular}
\end{center}
\caption{Semi-tractability index \(\Gamma\) of different algorithms for continuous time optimal stopping problems. \label{tab:cont}}
\end{table}

\section{Numerical experiments}

In the following experiments we illustrate the WSM algorithm in the case of
continuous time optimal stopping problems. Lower bounds for the WSM algorithm can be obtained
using a suboptimal policy computed on an independent set of trajectories. This policy can be constructed either directly via  \eqref{app} or by using   interpolation of the likelihood weights
\[
\frac{p(Z_{l+1}^{(j)}|\cdot)}{\sum_{m=1}^{N}p(Z_{l+1}%
^{(j)}|Z_{l}^{(m)})}.
\]
The fastest and simplest way to do this is to use the nearest neighbour
interpolation based on training set of trajectories, in all experiments below the number of
neighbours was set to $500.$

\subsection{\textbf{An American put on a single asset}}

\begin{figure}[tbh]
\centering
\begin{tabular}
[c]{cc}%
\includegraphics[scale=.3]{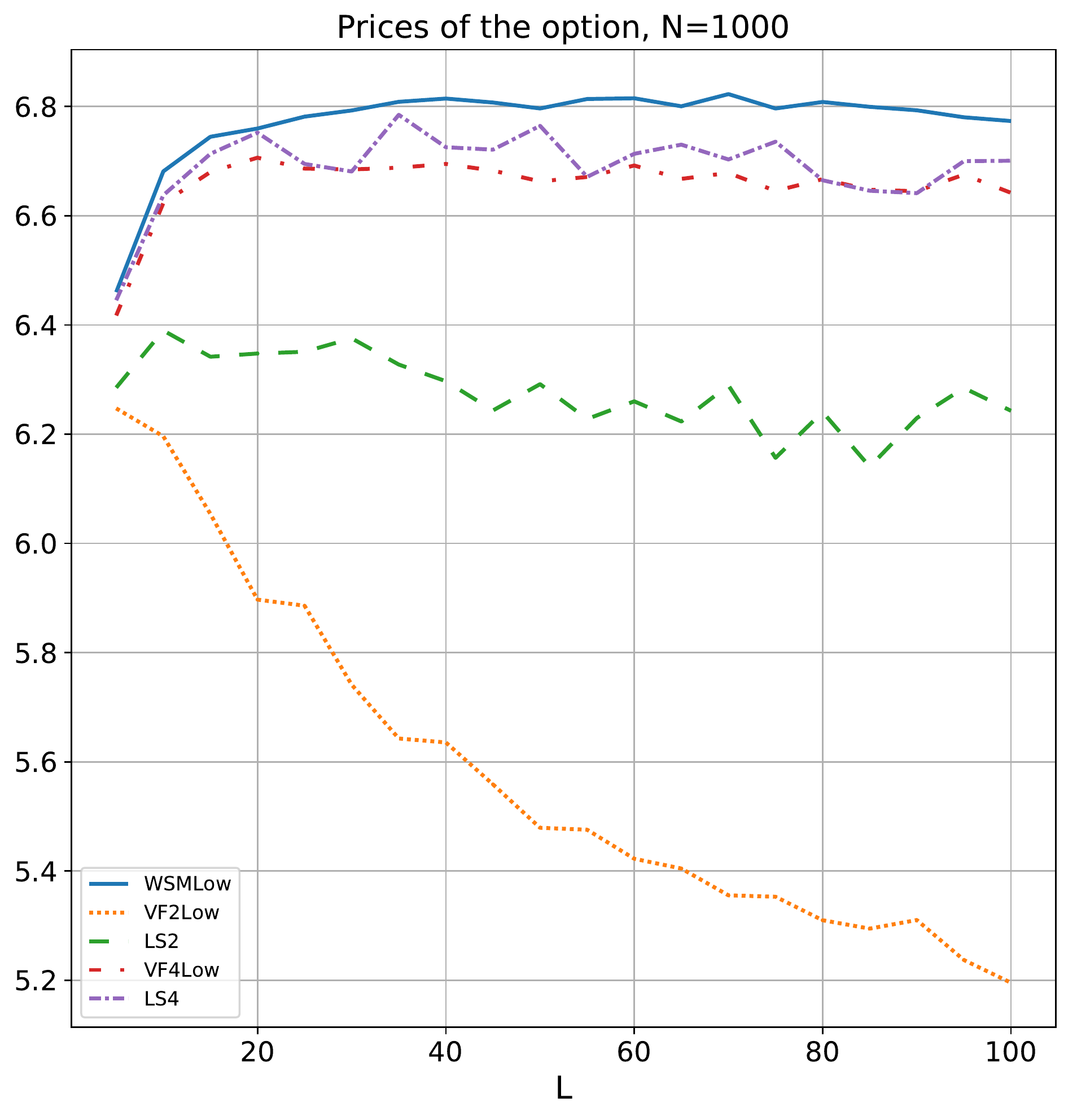} &
\includegraphics[scale=.3]{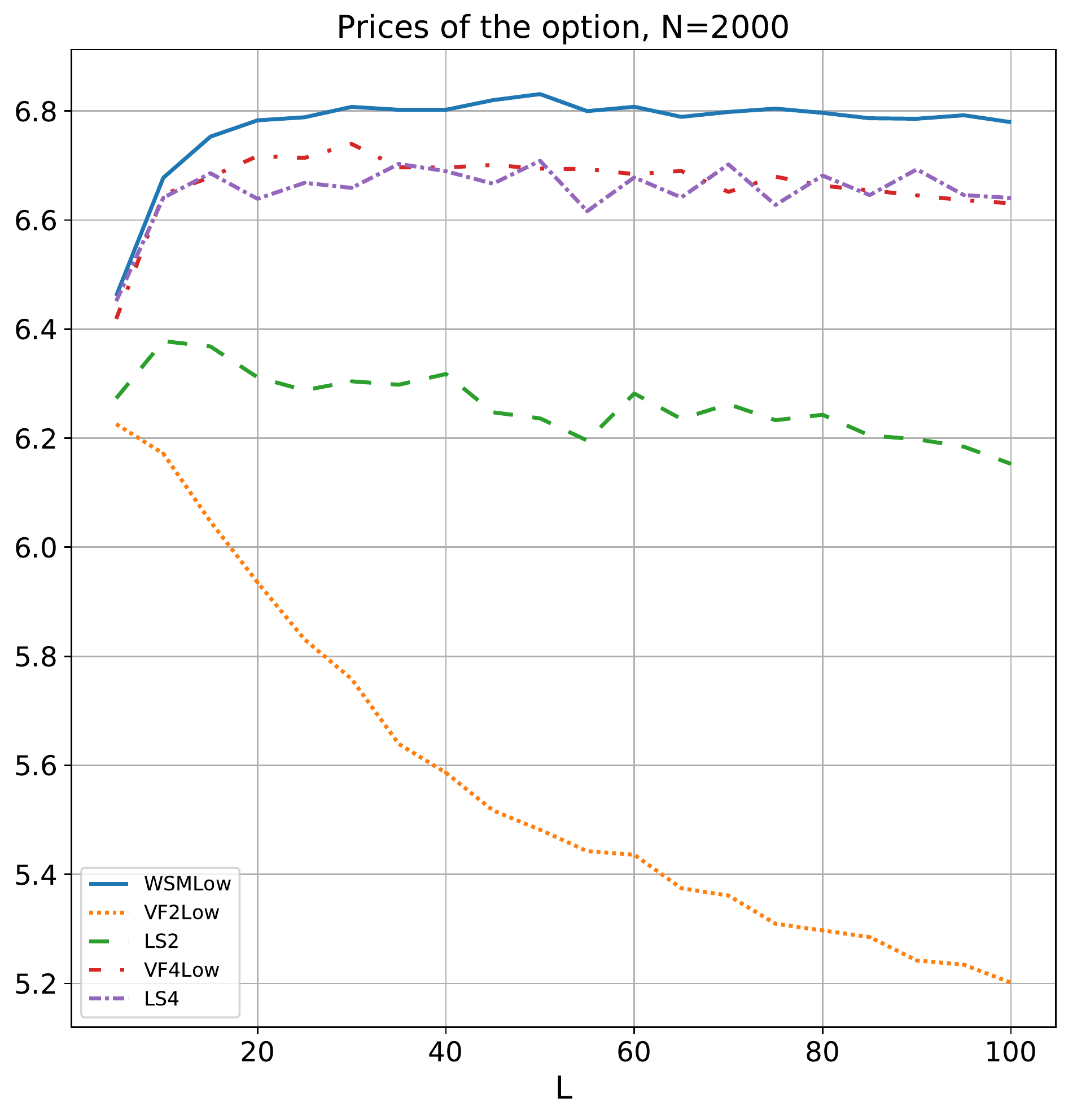}\\
(a) & (b)
\end{tabular}
\caption{Lower bounds for the price of a one-dimensional American put option
approximated using different methods and a uniform grid $t_{k}=kT/L,k=0,\ldots,L,$ of exercise dates.
 The numbers of training paths are $N_{\mathrm{{train}}}=1000$ (a) and $N_{\mathrm{{train}}}=2000$ (b),
and the number of new trajectories used to construct lower bounds is
$N_{\mathrm{{test}}}=20000$ in both cases. In LS and VF regression methods a polynomial basis of degree \(2\) and \(4\) is used.}%
\label{pic:put1d}%
\end{figure}

In order to illustrate the performance of  the WSM algorithm in continuous time,
we consider a financial problem of pricing American put option on a single
log-Brownian asset
\[
X_{t}=X_{0}\exp(\sigma W_{t}+(r-\sigma/2)t),
\]
with $r$ denoting the riskless rate of interest, assumed to be constant, and
$\sigma$ denoting the constant volatility. The payoff function is given by
$g(x)=(K-x)^{+}$ and a fair price of the option is given by
\[
U_{0}=\sup_{\tau\in\mathcal{T}[0,T]}\mathsf{E}\left[  e^{-r\tau}g(X_{\tau
})\right]  .
\]
No closed-form solution for the price of this option is known, but there are
various numerical methods which give accurate approximations to $V_{0}$. The
parameter values used are $r=0.08,$ $\sigma=0.20,$ $\delta=0,$ $K=100,$ $T=3$.
An accurate estimate for the true price obtained via a binomial tree type algorithm is
$6.9320$ (see \cite{BJK:D1putOption}). In Figure~\ref{pic:put1d} we show
lower bounds due to WSM, the least squares approaches of Longstaff and Schwartz~\cite{J_LS2001} (LS) and value function regression algorithm of  Tsitsiklis and Van Roy~\cite{J_TV2001} (VF) as functions of the
number of stopping times $L$ forming a uniform grid on $[0,T].$  These lower bounds  are constructed using a suboptimal stopping rule due to estimated  continuation values evaluated on a new independent set of trajectories.  The maximal degree of polynomials used as basis functions in LS and VF are indicated by the numbers (\(2\) and \(4\)) in legend. As can be
seen  WSM lower bounds are more stable when $L$  increases. The
VF lower bounds seem to diverge as $L\rightarrow\infty.$

\section{Proofs}

\label{sec:proofs}

\subsection{Proof of Proposition~\ref{eq:u-ut}}

For $l=L$ the statement reads%
\[
\int\left\vert U_{L}(x)-\widetilde{U}_{L}(x)\right\vert p_{L}(x|x_{0}%
)dx=\int1_{\left\vert x-x_{0}\right\vert >R}\,g(x)p_{L}(x|x_{0})dx=\varepsilon
_{L,R},
\]
so then it is true. Suppose (\ref{es}) is true for $0<l+1\leq L.$ Then, by
using $\left\vert \max(a,b)-\max(a,c)\right\vert \leq|b-c|$ and the fact that
$\widetilde{U}_{l}(x)$ vanishes for $\left\vert x-x_{0}\right\vert >R,$%
\begin{align*}
\left\vert U_{l}(x)-\widetilde{U}_{l}(x)\right\vert  &  \leq1_{\left\vert
x-x_{0}\right\vert \leq R}\left\vert \max\left[  g(x),\mathsf{E}\left[
\left.  U_{l+1}(X_{l+1})\right\vert X_{l}=x\right]  \right]  \right. \\
&  \left.  -\max\left[  g(x),\mathsf{E}\left[  \left.  \widetilde{U}%
_{l+1}(X_{l+1})\right\vert X_{l}=x\right]  \right]  \right\vert +1_{\left\vert
x-x_{0}\right\vert >R}U_{l}(x)\\
&  \leq1_{\left\vert x-x_{0}\right\vert \leq R}\mathsf{E}\left[  \left.
\left\vert U_{l+1}(X_{l+1})-\widetilde{U}_{l+1}(X_{l+1})\right\vert
\right\vert X_{l}=x\right]  +1_{\left\vert x-x_{0}\right\vert >R}U_{l}(x).
\end{align*}
Hence we have by induction,%
\begin{align*}
&  \int\left\vert U_{l}(x)-\widetilde{U}_{l}(x)\right\vert p_{l}(x|x_{0})dx\\
&  \leq\int1_{\left\vert x-x_{0}\right\vert >R}\mathsf{E}\left[  \left.
\left\vert U_{l+1}(X_{l+1})-\widetilde{U}_{l+1}(X_{l+1})\right\vert
\right\vert X_{l}=x\right]  p_{l}(x|x_{0})dx+\varepsilon_{l,R}\\
&  \leq\int\left\vert U_{l+1}(y)-\widetilde{U}_{l+1}(y)\right\vert
p_{l+1}(y|x_{0})dy+\varepsilon_{l,R}\\
&  =\sum_{j=l+1}^{L}\varepsilon_{j,R}+\varepsilon_{l,R}=\sum_{j=l}%
^{L}\varepsilon_{j,R}.
\end{align*}

\subsection{\bigskip Proof of Proposition~\ref{corL}}

Combining the assumptions (\ref{ag}) and (\ref{aX}) yields,
\begin{align*}
U_{l}(x)  &  =\operatornamewithlimits{esssup}_{\tau\in\mathcal{T}_{l,L}%
}\mathsf{E}\left[  \left.  g_{\tau}(Z_{\tau})\right\vert Z_{l}=x\right] \\
&  \leq c_{g}\mathsf{E}\left[  \left.  1+\max_{l\leq l^{\prime}\leq
L}\left\vert Z_{l^{\prime}}\right\vert \right\vert Z_{l}=x\right] \\
&  \leq c_{g}\left(  1+c_{Z}\right)  +c_{g}c_{Z}\left\vert x\right\vert .
\end{align*}
Using%
\begin{align*}
\int_{\left\vert x-x_{0}\right\vert >R}e^{-\frac{|x-x_{0}|^{2}}{2\alpha l}}dx
&  \leq e^{-\frac{R^{2}}{8\alpha l}}\left(  4/3\right)  ^{d/2}(2\pi\alpha
l)^{d/2},\text{ \ \ and}\\
\int_{\left\vert x-x_{0}\right\vert >R}\left\vert x-x_{0}\right\vert
e^{-\frac{|x-x_{0}|^{2}}{2\alpha l}}dx  &  \leq\sqrt{\int_{\left\vert
x-x_{0}\right\vert >R}e^{-\frac{|x-x_{0}|^{2}}{2\alpha l}}dx}\sqrt
{\int\left\vert x-x_{0}\right\vert ^{2}e^{-\frac{|x-x_{0}|^{2}}{2\alpha l}}%
dx}\\
&  \leq e^{-\frac{R^{2}}{8\alpha l}}2^{d/4}(2\pi\alpha l)^{d/2}\sqrt{d\alpha
l}%
\end{align*}
we get (note that $\left(  4/3\right)  ^{1/2}<2^{1/4}$),%
\begin{align*}
\varepsilon_{l,R}  &  \leq\frac{\varkappa}{(2\pi\alpha l)^{d/2}}%
\int_{\left\vert x-x_{0}\right\vert >R}\left(  c_{g}\left(  1+c_{Z}\right)
+c_{g}c_{Z}\left\vert x\right\vert \right)  e^{-\frac{|x-x_{0}|^{2}}{2\alpha
l}}dx\\
&  \leq\frac{\varkappa c_{g}\left(  1+c_{Z}+c_{Z}\left\vert x_{0}\right\vert
\right)  }{(2\pi\alpha l)^{d/2}}\int_{\left\vert x-x_{0}\right\vert
>R}e^{-\frac{|x-x_{0}|^{2}}{2\alpha l}}dx\\
&  +\frac{\varkappa c_{g}c_{Z}}{(2\pi\alpha l)^{d/2}}\int_{\left\vert
x-x_{0}\right\vert >R}\left\vert x-x_{0}\right\vert e^{-\frac{|x-x_{0}|^{2}%
}{2\alpha l}}dx\\
&  \leq\varkappa c_{g}\left(  1+c_{Z}+c_{Z}\left\vert x_{0}\right\vert
+c_{Z}\sqrt{d\alpha}\sqrt{l}\right)  2^{d/4}e^{-\frac{R^{2}}{8\alpha l}}\\
&  \equiv\left(  A+B\sqrt{l}\right)  c_{g}\varkappa e^{-\frac{R^{2}}{8\alpha
l}},
\end{align*}
for $l\geq1$ ($\varepsilon_{0,R}=0$ for $R>0$). Now by (\ref{es}), i.e.
Proposition~\ref{eq:u-ut}, we get%
\[
\int\bigl |U_{l}(x)-\widetilde{U}_{l}(x)\bigr |p_{l}(x|x_{0})\,dx\leq L\left(
A+B\sqrt{L}\right)  c_{g}\varkappa e^{-\frac{R^{2}}{8\alpha L}},
\]
whence the estimate (\ref{es2}).

\subsection{Proof of Proposition~\ref{thm: main-err}}

Let us write the sample based backward dynamic program (\ref{ABDP}) for step
$l<L$ in the form,%

\begin{equation}
\overline{U}_{l}\left(  Z_{l}^{(i)}\right)  =\mathbbm{1}_{\left\vert
Z_{l}^{(i)}-x_{0}\right\vert \leq R}\max\left[  g_{l}(Z_{l}^{(i)}),\sum
_{j=1}^{N}\overline{U}_{l+1}(Z_{l+1}^{(j)})\overline{w}_{ij}\right]
\label{wfo}%
\end{equation}
by defining the weights
\begin{equation}
w_{ij}:=\frac{p(Z_{l+1}^{(j)}|Z_{l}^{(i)})}{\sum_{m=1}^{N}p(Z_{l+1}%
^{(j)}|Z_{l}^{(m)})}, \label{wb}%
\end{equation}
where $l$ is fixed and suppressed. Let us further abbreviate%
\[
\mathcal{E}[f](x)=\mathsf{E}\left[  \left.  f(Z_{l+1})\right\vert
Z_{l}=x\right]  =\int f(y)p(y|x)dy
\]
for a generic Borel function $f\geq0.$ Using,
\[
\widetilde{U}_{l}\left(  Z_{l}^{(i)}\right)  =\mathbbm{1}_{\left\vert
Z_{l}^{(i)}-x_{0}\right\vert \leq R}\max\left[  g_{l}(Z_{l}^{(i)}%
),\mathcal{E}[\widetilde{U}_{l+1}](Z_{l}^{(i)})\right]  ,
\]
(\ref{wfo}), and $\left\vert \max(a,b)-\max(a,c)\right\vert \leq|b-c|,$ we
thus get
\begin{align}
&  \left\vert \overline{U}_{l}-\widetilde{U}_{l}\right\vert _{N}:=\frac{1}%
{N}\sum_{i=1}^{N}\left\vert \overline{U}_{l}(Z_{l}^{(i)})-\widetilde{U}%
_{l}(Z_{l}^{(i)})\right\vert \leq\nonumber\\
&  \frac{1}{N}\sum_{i=1}^{N}\mathbbm{1}_{\left\vert Z_{l}^{(i)}-x_{0}%
\right\vert \leq R}\left\vert \sum_{j=1}^{N}\overline{U}_{l+1}(Z_{l+1}%
^{(j)})w_{ij}-\mathcal{E}[\widetilde{U}_{l+1}](Z_{l}^{(i)})\right\vert
\nonumber\\
&  \leq\frac{1}{N}\sum_{i=1}^{N}\mathbbm{1}_{\left\vert Z_{l}^{(i)}%
-x_{0}\right\vert \leq R}\sum_{j=1}^{N}w_{ij}\left\vert \overline{U}%
_{l+1}(Z_{l+1}^{(j)})-\widetilde{U}_{l+1}(Z_{l+1}^{(j)})\right\vert
\nonumber\\
&  +\frac{1}{N}\sum_{i=1}^{N}\mathbbm{1}_{\left\vert Z_{l}^{(i)}%
-x_{0}\right\vert \leq R}\left\vert \sum_{j=1}^{N}\widetilde{U}_{l+1}%
(Z_{l+1}^{(j)})w_{ij}-\mathcal{E}[\widetilde{U}_{l+1}](Z_{l}^{(i)})\right\vert
\nonumber\\
&  \leq:\left\vert \overline{U}_{l+1}-\widetilde{U}_{l+1}\right\vert
_{N}+\mathcal{R}_{l+1}, \label{it}%
\end{align}
using that the weights in (\ref{wb}) sum up to one. One thus gets by iterating
(\ref{it}),%
\begin{equation}
\left\vert \overline{U}_{k}-\widetilde{U}_{k}\right\vert _{N}\leq\sum
_{l=k}^{L-1}R_{l+1} \label{it1}%
\end{equation}
since $\overline{U}_{L}-\widetilde{U}_{L}=0.$ Let us now introduce%
\begin{equation}
w_{ij}^{\circ}:=\frac{1}{N}\frac{p(Z_{l+1}^{(j)}|Z_{l}^{(i)})}{p_{l+1}%
(Z_{l+1}^{(j)}|x_{0})}, \label{wo}%
\end{equation}
and consider the generic term%
\begin{align*}
\mathcal{R}_{l+1}  &  =\frac{1}{N}\sum_{i=1}^{N}\mathbbm{1}_{\left\vert
Z_{l}^{(i)}-x_{0}\right\vert \leq R}\left\vert \sum_{j=1}^{N}\widetilde
{U}_{l+1}(Z_{l+1}^{(j)})w_{ij}-\mathcal{E}[\widetilde{U}_{l+1}](Z_{l}%
^{(i)})\right\vert \\
&  \leq\frac{1}{N}\sum_{i=1}^{N}\mathbbm{1}_{\left\vert Z_{l}^{(i)}%
-x_{0}\right\vert \leq R}\sum_{j=1}^{N}\widetilde{U}_{l+1}(Z_{l+1}%
^{(j)})\left\vert w_{ij}-w_{ij}^{\circ}\right\vert \\
&  +\frac{1}{N}\sum_{i=1}^{N}\mathbbm{1}_{\left\vert Z_{l}^{(i)}%
-x_{0}\right\vert \leq R}\left\vert \sum_{j=1}^{N}\left(  w_{ij}^{\circ
}\widetilde{U}_{l+1}(Z_{l+1}^{(j)})-\frac{1}{N}\mathcal{E}[\widetilde{U}%
_{l+1}](Z_{l}^{(i)})\right)  \right\vert \\
&  =:\text{Term}_{1}+\text{Term}_{2}.
\end{align*}
\ Due to (\ref{boundG}) one has,%
\[
\text{Term}_{1}\leq\frac{G_{R}}{N}\sum_{i=1}^{N}\sum_{j=1}^{N}%
\mathbbm{1}_{\left\vert Z_{l}^{(i)}-x_{0}\right\vert \leq R}%
\mathbbm{1}_{\left\vert Z_{l+1}^{(j)}-x_{0}\right\vert \leq R}\left\vert
w_{ij}-w_{ij}^{\circ}\right\vert ,
\]
and due to (\ref{wb}) and (\ref{wo}) we may write,%
\begin{align*}
\left\vert w_{ij}-w_{ij}^{\circ}\right\vert  &  =\left\vert \frac
{p(Z_{l+1}^{(j)}|Z_{l}^{(i)})}{\sum_{m=1}^{N}p(Z_{l+1}^{(j)}|Z_{l}^{(m)}%
)}-\frac{1}{N}\frac{p(Z_{l+1}^{(j)}|Z_{l}^{(i)})}{p_{l+1}(Z_{l+1}^{(j)}%
|x_{0})}\right\vert \\
&  =\frac{p(Z_{l+1}^{(j)}|Z_{l}^{(i)})}{\sum_{m=1}^{N}p(Z_{l+1}^{(j)}%
|Z_{l}^{(m)})}\left\vert 1-\frac{\frac{1}{N}\sum_{m=1}^{N}p(Z_{l+1}%
^{(j)}|Z_{l}^{(m)})}{p_{l+1}(Z_{l+1}^{(j)}|x_{0})\,}\right\vert .
\end{align*}
and so obtain,%
\begin{equation}
\text{Term}_{1}\leq\frac{G_{R}}{N}\sum_{j=1}^{N}\mathbbm{1}_{\left\vert
Z_{l+1}^{(j)}-x_{0}\right\vert \leq R}\left\vert 1-\frac{\frac{1}{N}\sum
_{m=1}^{N}p(Z_{l+1}^{(j)}|Z_{l}^{(m)})}{p_{l+1}(Z_{l+1}^{(j)}|x_{0}%
)}\right\vert .\nonumber
\end{equation}
We are now going to estimate
\[
\mathsf{E}\left[  \mathcal{R}_{l+1}\right]  \lesssim\mathsf{E}\left[
\text{Term}_{1}\right]  +\mathsf{E}\left[  \text{Term}_{2}\right]  .
\]
It holds that
\begin{multline*}
\mathsf{E}\left[  \text{Term}_{1}\right]  \leq\frac{G_{R}}{N}\mathsf{E}\left[
\mathbbm{1}_{\left\vert Z_{l+1}^{(1)}-x_{0}\right\vert \leq R}\left\vert
\sum_{m=1}^{N}\left(  1-\frac{p(Z_{l+1}^{(1)}|Z_{l}^{(m)})}{p_{l+1}%
(Z_{l+1}^{(1)}|x_{0})}\right)  \right\vert \right] \\
\leq\frac{G_{R}}{N}D_{R}+\frac{G_{R}}{N}\mathsf{E}\left[  \left\vert
\sum_{m=2}^{N}\mathbbm{1}_{\left\vert Z_{l+1}^{(1)}-x_{0}\right\vert \leq
R}\left(  1-\frac{p(Z_{l+1}^{(1)}|Z_{l}^{(m)})}{p_{l+1}(Z_{l+1}^{(1)}|x_{0}%
)}\right)  \right\vert \right]
\end{multline*}
with%
\[
D_{R}:=\mathsf{E}\left[  \mathbbm{1}_{\left\vert Z_{l+1}^{(1)}-x_{0}%
\right\vert \leq R}\left\vert 1-\frac{p(Z_{l+1}^{(1)}|Z_{l}^{(1)})}%
{p_{l+1}(Z_{l+1}^{(1)}|x_{0})}\right\vert \right]  .
\]
Now consider the i.i.d. random variables,%
\[
\eta_{m}^{(l+1)}:=\mathbbm{1}_{\left\vert Z_{l+1}^{(1)}-x_{0}\right\vert \leq
R}\left(  1-\frac{p(Z_{l+1}^{(1)}|Z_{l}^{(m)})}{p_{l+1}(Z_{l+1}^{(1)}|x_{0}%
)}\right)  ,\text{ \ \ }m=2,...,N,
\]
which have zero mean. Then, by Cauchy-Schwartz one has that%
\begin{align*}
\mathsf{E}\left\vert \sum_{m=2}^{N}\eta_{m}^{(l+1)}\right\vert  &  \leq
\sqrt{\mathsf{E}\left(  \sum_{m=2}^{N}\eta_{m}^{(l+1)}\right)  ^{2}}%
=E_{R}\sqrt{N}\text{ \ \ with}\\
E_{R}^{2}  &  :=\mathsf{Var}\left(  \eta_{2}^{(l+1)}\right)  =\mathsf{E}%
\left[  \mathbbm{1}_{\left\vert Z_{l+1}^{(1)}-x_{0}\right\vert \leq
R}\left\vert 1-\frac{p(Z_{l+1}^{(1)}|Z_{l}^{(2)})}{p_{l+1}(Z_{l+1}^{(1)}%
|x_{0})}\right\vert ^{2}\right]  ,
\end{align*}
Concerning Term$_{2}$, let us write%
\begin{align*}
\mathcal{E}[\widetilde{U}_{l+1}](Z_{l}^{(i)})  &  =\int\widetilde{U}%
_{l+1}(y)\frac{p(y|Z_{l}^{(i)})}{p_{l+1}(y|x_{0})}p_{l+1}(y|x_{0})dy\\
&  =\mathsf{E}\left[  \widetilde{U}_{l+1}(Z_{l+1}^{0,x_{0}})\frac
{p(Z_{l+1}^{0,x_{0}}|Z_{l}^{(i)})}{p_{l+1}(Z_{l+1}^{0,x_{0}}|x_{0})}\right]  ,
\end{align*}
where $Z^{0,x_{0}}$ is an independent dummy trajectory. We thus have%
\begin{multline*}
\mathsf{E}\left[  \text{Term}_{2}\right]  =\mathsf{E}\left[
\mathbbm{1}_{\left\vert Z_{l}^{(1)}-x_{0}\right\vert \leq R}\left\vert \left(
w_{11}^{\circ}\widetilde{U}_{l+1}(Z_{l+1}^{(1)})-\frac{1}{N}\mathcal{E}%
[\widetilde{U}_{l+1}](Z_{l}^{(1)})\right)  \right\vert \right] \\
+\mathsf{E}\left[  \left\vert \sum_{j=2}^{N}\zeta_{j}^{(l+1)}\right\vert
\right]  ,
\end{multline*}
where for $j=2,...,N,$ the random variables%
\begin{multline*}
\zeta_{j}^{(l+1)}:=\mathbbm{1}_{\left\vert Z_{l}^{(1)}-x_{0}\right\vert \leq
R}\left(  w_{1j}^{\circ}\widetilde{U}_{l+1}(Z_{l+1}^{(j)})-\frac{1}%
{N}\mathcal{E}[\widetilde{U}_{l+1}](Z_{l}^{(1)})\right) \\
=\frac{\mathbbm{1}_{\left\vert Z_{l}^{(1)}-x_{0}\right\vert \leq R}}{N}\left(
\frac{p(Z_{l+1}^{(j)}|Z_{l}^{(1)})}{p_{l+1}(Z_{l+1}^{(j)}|x_{0})}\widetilde
{U}_{l+1}(Z_{l+1}^{(j)})-\mathsf{E}\left[  \widetilde{U}_{l+1}(Z_{l+1}%
^{0,x_{0}})\frac{p(Z_{l+1}^{0,x_{0}}|Z_{l}^{(1)})}{p_{l+1}(Z_{l+1}^{0,x_{0}%
}|x_{0})}\right]  \right)
\end{multline*}
are i.i.d. and have zero mean. We so have by Cauchy-Schwartz again,%
\begin{align*}
\mathsf{E}\left[  \left\vert \sum_{j=2}^{N}\zeta_{j}^{(l+1)}\right\vert
\right]   &  \leq\sqrt{\mathsf{E}\left(  \sum_{j=2}^{N}\zeta_{j}%
^{(l+1)}\right)  ^{2}}=\sqrt{N\mathsf{Var}\left(  \zeta_{2}^{(l+1)}\right)
}\leq F_{R}G_{R}/\sqrt{N},\text{ \ \ where}\\
F_{R}^{2}  &  =\mathsf{E}\left[  \mathbbm{1}_{\left\vert Z_{l}^{(1)}%
-x_{0}\right\vert \leq R}\left\vert \frac{p(Z_{l+1}^{(2)}|Z_{l}^{(1)}%
)}{p_{l+1}(Z_{l+1}^{(2)}|x_{0})}\right\vert ^{2}\right] \\
&  =\int\int_{|y-x_{0}|\leq R}\frac{p^{2}(y|x)}{p_{l+1}(y|x_{0})}p_{l}%
(x|x_{0})\,dxdy.
\end{align*}
Secondly, one has%
\begin{align*}
&  \mathsf{E}\left[  \mathbbm{1}_{\left\vert Z_{l}^{(1)}-x_{0}\right\vert \leq
R}\left\vert \left(  w_{11}^{\circ}\widetilde{U}_{l+1}(Z_{l+1}^{(1)})-\frac
{1}{N}\mathcal{E}[\widetilde{U}_{l+1}](Z_{l}^{(1)})\right)  \right\vert
\right] \\
&  \leq\frac{1}{N}\mathsf{E}\left[  \mathbbm{1}_{\left\vert Z_{l}^{(1)}%
-x_{0}\right\vert \leq R}\frac{p(Z_{l+1}^{(1)}|Z_{l}^{(1)})}{p_{l+1}%
(Z_{l+1}^{(1)}|x_{0})}\widetilde{U}_{l+1}(Z_{l+1}^{(1)})\right] \\
&  +\frac{1}{N}\mathsf{E}\left[  \mathbbm{1}_{\left\vert Z_{l}^{(1)}%
-x_{0}\right\vert \leq R}\mathsf{E}\left[  \widetilde{U}_{l+1}(Z_{l+1}%
^{0,x_{0}})\frac{p(Z_{l+1}^{0,x_{0}}|Z_{l}^{(1)})}{p_{l+1}(Z_{l+1}^{0,x_{0}%
}|x_{0})}\right]  \right] \\
&  \leq\frac{G_{R}}{N}\mathsf{E}\left[  \mathbbm{1}_{\left\vert Z_{l+1}%
^{(1)}-x_{0}\right\vert \leq R}\frac{p(Z_{l+1}^{(1)}|Z_{l}^{(1)})}%
{p_{l+1}(Z_{l+1}^{(1)}|x_{0})}\right] \\
&  +\frac{G_{R}}{N}\mathsf{E}\left[  \mathbbm{1}_{\left\vert Z_{l+1}^{0,x_{0}%
}-x_{0}\right\vert \leq R}\frac{p(Z_{l+1}^{0,x_{0}}|Z_{l}^{(1)})}%
{p_{l+1}(Z_{l+1}^{0,x_{0}}|x_{0})}\right] \\
&  =:\frac{G_{R}}{N}H_{R}.
\end{align*}
Next it follows that%
\begin{align*}
D_{R}  &  \leq1+\mathsf{E}\left[  \mathbbm{1}_{\left\vert Z_{l+1}^{(1)}%
-x_{0}\right\vert \leq R}\frac{p(Z_{l+1}^{(1)}|Z_{l}^{(1)})}{p_{l+1}%
(Z_{l+1}^{(1)}|x_{0})}\right] \\
&  =1+\int p_{l}(x|x_{0})\,dx\int_{|y-x_{0}|\leq R}\frac{p^{2}(y|x)}%
{p_{l+1}(y|x_{0})}dy\\
&  \leq1+F_{R}^{2}.
\end{align*}
Further, one obviously has that $E_{R}^{2}\leq2+2F_{R}^{2},$ and $H_{R}%
\leq1+F_{R}^{2}$ since%
\[
\mathsf{E}\left[  \mathbbm{1}_{\left\vert Z_{l+1}^{0,x_{0}}-x_{0}\right\vert
\leq R}\frac{p(Z_{l+1}^{0,x_{0}}|Z_{l}^{(1)})}{p_{l+1}(Z_{l+1}^{0,x_{0}}%
|x_{0})}\right]  \leq1.
\]
By now taking the expectation in (\ref{it1}) and gathering all together we
obtain,%
\begin{align}
\mathsf{E}\left[  \bigl |\overline{U}_{k}-\widetilde{U}_{k}\bigr |_{N}\right]
&  \leq(L-k)G_{R}\left(  \frac{\sqrt{2+2F_{R}^{2}}+F_{R}}{\sqrt{N}}%
+\frac{2+2F_{R}^{2}}{N}\right) \label{gath}\\
&  \leq\left(  3+\sqrt{2}\right)  (L-k)G_{R}\frac{1+F_{R}}{\sqrt{N}},\nonumber
\end{align}
assuming that $N$ is taken such that $(1+F_{R})/\sqrt{N}<1.$\bigskip
\subsection{Proof of Proposition~\ref{compl-wsm-discr}}
In order to achieve a required accuracy $\varepsilon>0,$ let us take $R$ and
$N$ large enough such that both error terms in (\ref{errL}) are equal to
$\varepsilon/2.$ Hence, we first take%
\[
R_{\varepsilon,d}=\left(  8\alpha L\right)  ^{1/2}\log^{1/2}\frac
{Lc_{g}\varkappa\left(  1+c_{Z}+c_{Z}\left\vert x_{0}\right\vert +c_{Z}%
\sqrt{d\alpha L}\right)  2^{1+d/4}}{\varepsilon},
\]
that is $R\nearrow\infty$ when $d+\varepsilon^{-1}\nearrow\infty.$ Then take,
with $\asymp$ denoting asymptotic equivalence for $R\nearrow\infty$ up to some
natural constant,%
\begin{multline*}
N_{\varepsilon}\asymp L^{2}c_{g}^{2}\varkappa\left(  e/\alpha\right)
^{d/2}d^{-d/2}R_{\varepsilon}^{d+2}\varepsilon^{-2}\asymp\alpha c_{g}%
^{2}\varkappa\left(  8e/d\right)  ^{d/2}L^{d/2+3}\\
\times\varepsilon^{-2}\log^{d/2+1}\frac{L\left(  1+c_{Z}+c_{Z}\left\vert
x_{0}\right\vert +c_{Z}\sqrt{d\alpha L}\right)  2^{1+d/4}c_{g}\varkappa
}{\varepsilon}.
\end{multline*}
Thus, the computational work load (complexity) is given by%
\begin{multline}
c_{f}^{(d)}N_{\varepsilon}^{2}L\leq c_{1}\alpha^{2}c_{g}^{4}\varkappa^{2}%
c_{f}^{(d)}\left(  8e/d\right)  ^{d}L^{d+7}\label{comL}\\
\times\varepsilon^{-4}\log^{d+2}\frac{L\left(  1+c_{Z}+c_{Z}\left\vert
x_{0}\right\vert +c_{Z}\sqrt{d\alpha L}\right)  2^{1+d/4}c_{g}\varkappa
}{\varepsilon}%
\end{multline}
where $c_{1}$ is a natural constant. Now let us write%
\begin{multline*}
d^{-d}\log^{d+2}\frac{L\left(  1+c_{Z}+c_{Z}\left\vert x_{0}\right\vert
+c_{Z}\sqrt{d\alpha L}\right)  2^{1+d/4}c_{g}\varkappa}{\varepsilon}\\
=d^{2}\log^{d+2}\left[  \frac{L^{1/d}\left(  1+c_{Z}+c_{Z}\left\vert
x_{0}\right\vert +c_{Z}\sqrt{d\alpha L}\right)  ^{1/d}2^{1/d+1/4}\left(
c_{g}\varkappa\right)  ^{1/d}}{\varepsilon^{1/d}}\right]  .
\end{multline*}
Then, using the elementary estimate $\left(  a+b\sqrt{d}\right)  ^{1/d}\leq
ae^{b/a},$ for $a,b>0,$ $d\geq1,$ and assuming that $\varepsilon<1,$
(\ref{comL}) implies (\ref{comLL}).

\subsection{Proof of Proposition~\ref{errcirc}}
On the one hand one has%
\begin{align*}
U_{t_{l}}^{\circ}(X_{t_{l}})-U_{t_{l}}(\overline{X}_{t_{l}})  &
=\operatornamewithlimits{esssup}_{\tau\in\mathcal{T}_{l,L}}\mathsf{E}%
_{\mathcal{F}_{t_{l}}}\left[  g(\tau,X_{\tau})\right]
-\operatornamewithlimits{esssup}_{\overline{\tau}\in\mathcal{T}_{l,L}%
}\mathsf{E}_{\mathcal{F}_{t_{l}}}\left[  g(\overline{\tau},\overline
{X}_{\overline{\tau}})\right] \\
&  \leq\operatornamewithlimits{esssup}_{\tau\in\mathcal{T}_{l,L}}%
\mathsf{E}_{\mathcal{F}_{t_{l}}}\left[  g(\tau,X_{\tau})-g(\tau,\overline
{X}_{\tau})\right] \\
&  \leq\operatornamewithlimits{esssup}_{\tau\in\mathcal{T}_{l,L}}%
\mathsf{E}_{\mathcal{F}_{t_{l}}}\left[  \left\vert g(\tau,X_{\tau}%
)-g(\tau,\overline{X}_{\tau})\right\vert \right]  ,
\end{align*}
and on the other one has similarly%
\begin{align*}
U_{t_{l}}(\overline{X}_{t_{l}})-U_{t_{l}}^{\circ}(X_{t_{l}})  &
=\operatornamewithlimits{esssup}_{\overline{\tau}\in\mathcal{T}_{l,L}%
}\mathsf{E}_{\mathcal{F}_{t_{l}}}\left[  g(\overline{\tau},\overline
{X}_{\overline{\tau}})\right]  -\operatornamewithlimits{esssup}_{\tau
\in\mathcal{T}_{l,L}}\mathsf{E}_{\mathcal{F}_{t_{l}}}\left[  g(\tau,X_{\tau
})\right] \\
&  \leq\operatornamewithlimits{esssup}_{\overline{\tau}\in\mathcal{T}_{l,L}%
}\mathsf{E}_{\mathcal{F}_{t_{l}}}\left[  g(\overline{\tau},\overline
{X}_{\overline{\tau}})-g(\tau,X_{\tau})\right] \\
&  \leq\operatornamewithlimits{esssup}_{\tau\in\mathcal{T}_{l,L}}%
\mathsf{E}_{\mathcal{F}_{t_{l}}}\left[  \left\vert g(\tau,X_{\tau}%
)-g(\tau,\overline{X}_{\tau})\right\vert \right].
\end{align*}
Hence we get%
\begin{align*}
\mathsf{E}\left[  \left\vert U_{t_{l}}^{\circ}(X_{t_{l}})-U_{t_{l}}%
(\overline{X}_{t_{l}})\right\vert \right]   &  \leq\mathsf{E}\left[
\sup_{0\leq s\leq T}\left\vert g(s,X_{s})-g(s,\overline{X}_{s})\right\vert
\right] \\
&  \leq L_{g}\mathsf{E}\left[  \sup_{0\leq s\leq T}\left\vert X_{s}%
-\overline{X}_{s}\right\vert \right]  \leq C^{\text{Euler}}\sqrt{h},
\end{align*}
due to the strong order of the Euler scheme, with $L_{g}$ being some Lipschitz
constant for $g.$

\bibliographystyle{plain}
\bibliography{tractability}

\begin{thebibliography}{10}

\bibitem{agarwal2013comparing}
Ankush Agarwal and Sandeep Juneja.
\newblock Comparing optimal convergence rate of stochastic mesh and least
  squares method for bermudan option pricing.
\newblock In {\em Proceedings of the 2013 Winter Simulation Conference:
  Simulation: Making Decisions in a Complex World}, pages 701--712. IEEE Press,
  2013.

\bibitem{Azencott}
Robert Azencott.
\newblock Densit\'{e} des diffusions en temps petit: d\'{e}veloppements
  asymptotiques. {I}.
\newblock In {\em Seminar on probability, {XVIII}}, volume 1059 of {\em Lecture
  Notes in Math.}, pages 402--498. Springer, Berlin, 1984.

\bibitem{Bally}
Vlad Bally, Gilles Pag\`es, and Jacques Printems.
\newblock A quantization tree method for pricing and hedging multidimensional
  {A}merican options.
\newblock {\em Math. Finance}, 15(1):119--168, 2005.

\bibitem{belomestny2018advanced}
Denis Belomestny and John Schoenmakers.
\newblock {\em Advanced Simulation-Based Methods for Optimal Stopping and
  Control: With Applications in Finance}.
\newblock Springer, 2018.

\bibitem{J_BrGl}
M.~Broadie and P.~Glasserman.
\newblock A stochastic mesh method for pricing high-dimensional {A}merican
  options.
\newblock {\em Journal of Computational Finance}, 7(4):35--72, 2004.

\bibitem{CLP}
Emmanuelle Cl{\'e}ment, Damien Lamberton, and Philip Protter.
\newblock An analysis of a least squares regression method for american option
  pricing.
\newblock {\em Finance and Stochastics}, 6(4):449--471, 2002.

\bibitem{MR872464}
D.~Dacunha-Castelle and D.~Florens-Zmirou.
\newblock Estimation of the coefficients of a diffusion from discrete
  observations.
\newblock {\em Stochastics}, 19(4):263--284, 1986.

\bibitem{Florens-Zmirou}
Dani\'{e}le Florens-Zmirou.
\newblock On estimating the diffusion coefficient from discrete observations.
\newblock {\em J. Appl. Probab.}, 30(4):790--804, 1993.

\bibitem{goldberg2018beating}
David~A Goldberg and Yilun Chen.
\newblock Beating the curse of dimensionality in options pricing and optimal
  stopping.
\newblock {\em arXiv preprint arXiv:1807.02227}, 2018.

\bibitem{jaillet1990variational}
Patrick Jaillet, Damien Lamberton, and Bernard Lapeyre.
\newblock Variational inequalities and the pricing of american options.
\newblock {\em Acta Applicandae Mathematica}, 21(3):263--289, 1990.

\bibitem{BJK:D1putOption}
Beom Jin~Kim, Yong-Ki Ma, and Hi~Choe.
\newblock A simple numerical method for pricing an american put option.
\newblock {\em Journal of Applied Mathematics}, 2013, 02 2013.

\bibitem{Li}
Chenxu Li.
\newblock Maximum-likelihood estimation for diffusion processes via closed-form
  density expansions.
\newblock {\em Ann. Statist.}, 41(3):1350--1380, 2013.

\bibitem{J_LS2001}
F.A. Longstaff and E.S. Schwartz.
\newblock Valuing american options by simulation: a simple least-squares
  approach.
\newblock {\em Review of Financial Studies}, 14(1):113--147, 2001.

\bibitem{NoWo}
Erich Novak and Henryk Wo\'{z}niakowski.
\newblock {\em Tractability of multivariate problems. {V}ol. 1: {L}inear
  information}, volume~6 of {\em EMS Tracts in Mathematics}.
\newblock European Mathematical Society (EMS), Z\"{u}rich, 2008.

\bibitem{rust1997using}
John Rust.
\newblock Using randomization to break the curse of dimensionality.
\newblock {\em Econometrica: Journal of the Econometric Society}, pages
  487--516, 1997.

\bibitem{J_TV2001}
J.~Tsitsiklis and B.~Van~Roy.
\newblock Regression methods for pricing complex american style options.
\newblock {\em IEEE Trans. Neural. Net.}, 12(14):694--703, 2001.

\bibitem{Z}
Daniel~Z Zanger.
\newblock Quantitative error estimates for a least-squares monte carlo
  algorithm for american option pricing.
\newblock {\em Finance and Stochastics}, 17(3):503--534, 2013.

\end{thebibliography}

\end{document}